# Recent advances with THGEM detectors

**S. Bressler[*], L. Arazi, L. Moleri, M. Pitt, A. Rubin and A. Breskin**
*Department of Particle Physics and Astrophysics*
*Weizmann Institute of Science, 76100 Rehovot, Israel*
*E-mail*: shikma.bressler@cern.ch

ABSTRACT: The Thick Gaseous Electron Multiplier (THGEM) is a simple and robust electrode suitable for large area detectors. In this work the results of extensive comparative studies of the physical properties of different THGEM-based structures are reviewed. The focus is on newly suggested THGEM-like WELL configurations as well as on recently developed characterization methods. The WELL structures are single-sided THGEM electrodes directly coupled to different anode readout electrodes. The structures differ by the coupling concept of the bottom THGEM electrode to the metallic readout pads: a Thick WELL (THWELL) with direct coupling; the Resistive WELL (RWELL) and the Segmented Resistive WELL (SRWELL) coupled through thin resistive films deposited on insulating sheets and a Resistive-Plate WELL (RPWELL) coupled through a plate of high bulk resistivity. The results are compared to that of traditional double-sided THGEM electrodes followed by induction gaps - in some cases with moderate additional multiplication within the gap. We compare the different configurations in terms of gain, avalanche extension, discharge-rate and magnitude as well as rate capabilities over a broad dynamic range – exploiting a method that mimics highly ionizing particles in the laboratory. We report on recent studies of avalanche distribution in THGEM holes using optical readout.

# Contents

## 1. Introduction

The Thick Gaseous Electron Multiplier (THGEM) [1] has been gaining interest as a robust detector element. The operation and properties of this hole-multiplier, which can be mass-produced over large areas by the printed-circuit board industry, were systematically studied by many authors. The results with various detector configurations, gas mixtures, pressures and temperatures were reported in [2-8] and reviewed in [9]; they relate to single-element or to cascaded-elements of standard double-sided THGEM electrodes followed by an induction gap and a readout anode.

    THGEM multipliers are suitable for a variety of applications that do not require high spatial resolution. Current examples are CsI-coated cascaded-THGEM UV-photon detectors [10] for Ring-Imaging Cherenkov (RICH) detectors, replacing advantageously wire-chambers [11-15]; cryogenic gaseous photomultipliers for recording scintillation-light in noble-liquid detectors, developed for future dark-matter experiments, medical imaging cameras [16-20] and in combined neutron/gamma inspection systems [21]; fast-neutron detectors with dedicated converter-foils [22, 23] and thin THGEM-based sampling elements, developed mainly for digital hadronic calorimeters (DHCAL), in view of their potential deployment in high-energy physics experiments in future linear colliders [24-26].

    Over the past two years we have concentrated on the development of novel THGEM-like multipliers. We focused on thin, single-stage, configurations (described in detail below) in which stable operation conditions (with low discharge probability) are usually hard to achieve.

We thoroughly investigated the concept of a 'WELL' configuration [24, 25] - a THGEM with a closed bottom anode, similar to the "blind THGEM" [27]; in an attempt to reach efficient protection against occasional discharges – we investigated various sorts of resistive anodes, in different detector configurations.

Some methods were developed for investigating and evaluating the detectors' response. The techniques and the results of extensive comparative studies of the different detector structures are summarized here.

The different THGEM structures investigated are presented in section 2; the evaluation methods are discussed in section 3; the properties of the different structures are discussed in sections 4, followed by a summary in section 5.

## 2. The THGEM structures

The different THGEM structure investigated can be divided into two main categories: double-sided THGEM electrode configurations and single-sided WELL- configurations. Some multi-stage configurations comprising cascaded multipliers of two (or more) double-sided THGEMs or a WELL preceded by a double-sided THGEM were also investigated.

### 2.1 Double-sided THGEM structures

The most studied THGEM-based structures are double-sided electrodes copper-clad on both sides of the insulating material [1, 9]. In the typical configuration (Figure 1a), the THGEM electrode is preceded by a conversion and drift gap. Radiation-induced ionization electrons drift into the THGEM holes, where multiplication occurs under a high electric field. The resulting avalanche electrons are extracted into a few-mm wide induction gap and drift towards the readout anode, inducing a signal. As long as the field strength in the induction gap, $E_{induction}$, is smaller than the multiplication threshold (~2.5 kV/cm in Ne/5%CH$_4$), the avalanche-induced signal is characterized by a fast rise time (of a few ns); when a discharge occurs, its energy is often shared between the bottom THGEM electrode and the anode.

When $E_{induction}$ exceeds the multiplication threshold, the signal is characterized by a fast rise time followed by a slow ion component – resulting of the additional multiplication within the induction gap [28, 29]. In such mode of operation, easily reached with a narrow induction gap (relatively small applied voltage), a large fraction of the discharge energy reaches the anode [28] – with a higher potential to damage the readout electronics.

### 2.2 Single-sided WELL structures

The Thick-WELL (THWELL), shown in Figure 1b, was first suggested in [24] (a similar idea was presented in [27]). It consists of a single-sided THGEM electrode (copper-clad on its top side only) coupled directly to the anode. The absence of the induction gap leads to a significantly thinner geometry. Compared to a THGEM with an induction gap, higher gains could be obtained for lower applied voltage across the THGEM electrode, due to the larger electric field within the closed holes [24]. Similar to a double-sided configuration operated with multiplication in the induction gap, the signal is characterized by a fast rise (the avalanche electrons) followed by a slow avalanche-ion component. In case of a discharge all the resulting energy reaches the anode.

The resistive-WELL (RWELL), Figure 1c, and the segmented resistive-WELL (SRWELL), Figure 1d, described in [24-26], were proposed in an attempt to protect the multiplier, its anode and the readout electronics from discharge damages and to minimise eventual dead-time effects following a discharge. In an RWELL configuration (Figure 1c), the THGEM WELL-electrode is coupled directly to a resistive anode; the resistive layer is deposited on a 0.1 mm thick FR4 sheet (using a spraying technique described in [30]), in contact with the metal readout pads. The avalanche inductive signal is recorded on the anode, similar in shape to that of the THWELL [24]. The resistive layer (with surface resistivity in the range between 1-20 MΩ/square) serves to effectively quench the energy of occasional discharges. However, charge spreading across the resistive layer surface was found to cause considerable inter-pad cross-talk [24].

In the SRWELL (Figure 1d), the charge spreading across the resistive layer is stopped at the pad border by adding a grid of thin copper lines below the resistive layer, matching the pad dimensions – of $1 \times 1$ cm$^2$ in our detector prototypes (Figure 2); the grid allows for rapid clearance of the electrons diffusing over its surface. Similarly, the holes of the SRWELL are drilled in a square pattern, with 'blind' copper strips above the pad boundaries; these were designed to prevent discharges in holes residing directly above the metal grid lines [25].

In the most recent WELL-structure, the Resistive-Plate WELL (RPWELL) [31] (Figure 1e), the WELL electrode is coupled to a thin plate of high bulk resistivity ($\sim 10^9 - 10^{12}$ Ωcm), with the metal anode pads placed behind the plate. Compared to the RWELL, this geometry has higher resistivity which should provide superior spark protection and quenching. In addition, lateral charge spreading should be suppressed since the accumulated charge is transported through the layer (as opposed to transversely across its surface in the RWELL). The pulse shape on the anode is similar to that recorded on the other WELL configurations described above.

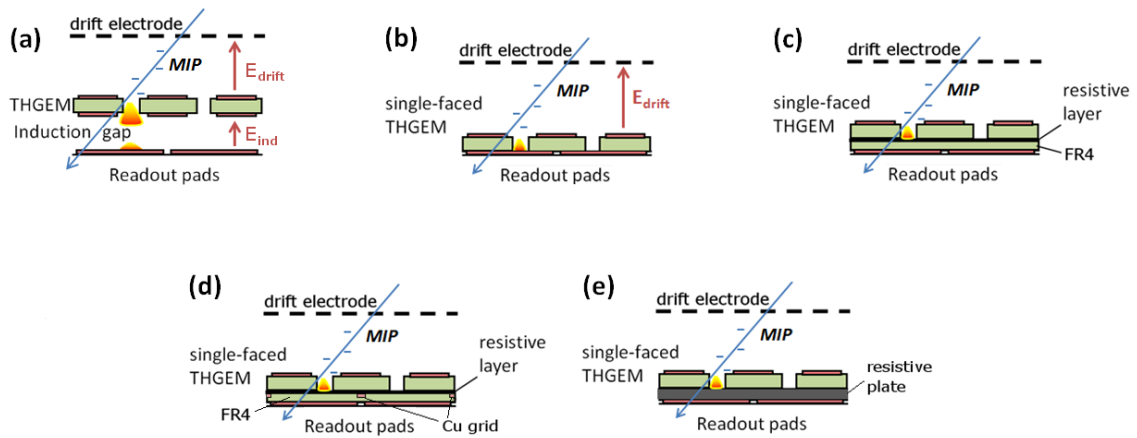

**Figure 1:** THGEM configurations: a) Double-sided THGEM with an induction gap; it usually transfers avalanche electrons to the anode but can also operate at additional moderate multiplication, to enhance the total gain b) THWELL: a single-sided THGEM electrode coupled to a metallic pad-readout anode. c) RWELL: a single sided THGEM coupled to the pad-readout anode through a thin resistive layer on a 100 μm thick insulator sheet d) SRWELL: similar to the RWELL but with the resistive layer deposited on the insulator - on top a conductive grid (see Figure 2) - minimizing the charge sharing between neighboring pads. e) RPWELL: A single-sided THGEM electrode coupled to metal pads through a plate of large bulk resistivity. In all configurations the avalanche-induced signal is recorded from the readout (anode) pads.

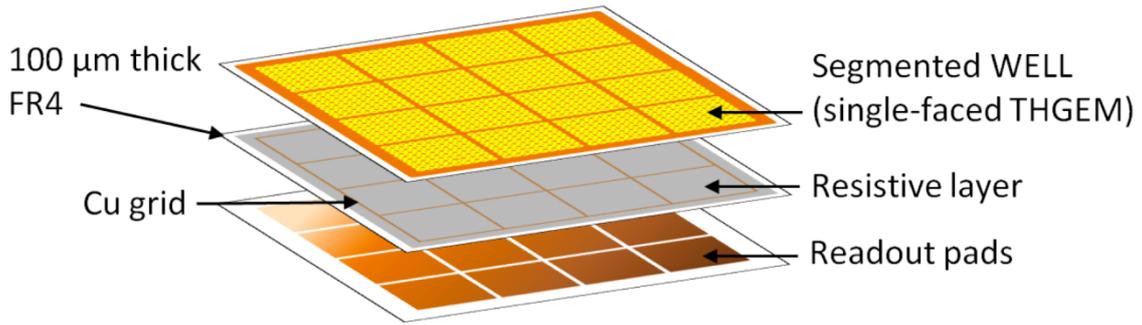

**Figure 2:** A schematic description of the SRWELL structure. The pad-readout anode (e.g. with 1 x 1 cm$^2$ pads) is followed by a thin (100 μm) FR4 layer. The copper grid on top of the FR4 follows the pad boundaries. A resistive layer is sprayed on top of the grid. The holes of the SRWELL are drilled in a square pattern, with 'blind' copper strips above the pad boundaries.

### 2.3 Multi-stage configurations

Multi-stage configurations comprising several double-sided THGEM electrodes in cascade were described in [9]. Compared to single-stage configurations, higher maximal achievable gains and lower discharge probabilities were recorded.

Double-stage configurations with a double-sided THGEM preceding an SRWELL were studied at CERN-SPS test beam facility, with a 100×100 mm$^2$ detector in Ne/5%CH$_4$, using 150 GeV/c muon and pion beams [25,26]. The total detector thickness was ~6 mm. Signals recorded with APV-SRS electronics [32] yielded detection efficiency values > 98%, with pad multiplicity (number of fired 1×1 cm$^2$ pads per track) of ~1.1; the discharge probability was low (~10$^{-7}$ for muons, ~10$^{-6}$ for pions. These results are compatible with the required performance of DHCAL sampling elements in the SiD detector planned for the future ILC or CLIC [33,34]. Details are provided in [26].

### 3. Evaluation methodologies

Two methods were employed to study the properties of our THGEM structures; both can be naturally applied to other gaseous radiation-detector concepts.

In our experiments, the investigated detectors were assembled in an aluminium chamber and continuously flushed with 1 atm of Ne/5%CH$_4$ (in some cases, indicated below, with Ne/5%CF$_4$). The detectors were irradiated with x-ray photons (of either $^{55}$Fe source or an x-ray tube) through a thin Kapton or Mylar window. The x-ray beams were usually collimated; the x-ray tube beam had additional nickel and copper filters to improve the spectrum.

When the detectors were operated at high gains with the signals well above noise, the anode signals were recorded with a charge-sensitive preamplifier followed by a linear amplifier and a multi-channel analyzer; the entire readout chain was charge-calibrated. At low-gain operation, with the signals buried within noise, the gain was estimated from measurements of the current flow through the anode; in these experiments, the measurements were calibrated to account for leakage currents and for the x-ray event rate.

In parallel, the currents provided by the power supplies to the different detector electrodes and the measured voltage values were carefully monitored (using an NI-DAQ ADC card type NI-USB 6008). In discharge-rate measurements, a discharge was defined as a rapid increase in the current supplied to two or more electrodes. Details are provided in [28].

## 3.1 The charge-injector: evaluation over a broad dynamic-range

A charge-injector concept was developed and applied to permit systematic investigations of detector prototypes in laboratory conditions mimicking highly ionizing events [28]. In the charge-injector setup (Figure 3), an additional THGEM multiplier ('injector') was mounted in front of the investigated detector. The injector multiplies the primary ionization electrons released by the incoming radiation within the drift gap, located between the cathode and the injector's top electrode; the multiplied electron swarm, whose size is tuned by the injector's voltage, is drifted through a transfer gap to the investigated detector. A transfer gap of 4-5 mm was found adequate for decoupling the investigated detector from the injector field [28].

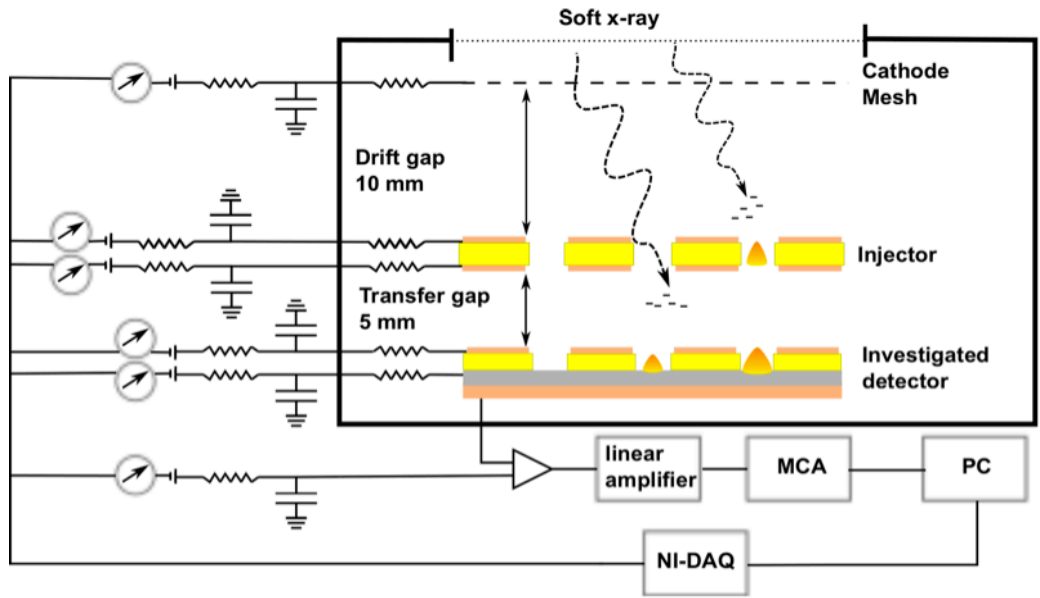

**Figure 3:** Schematic description of the charge-injector method. An investigated detector (here an RWELL) is preceded by a charge pre-amplifier, "injector" (here a THGEM). Impinging radiation (here soft x-rays) is converted within both a conversion/drift gap and a transfer gap between the two elements. X-rays converted in the drift gap above the THGEM are pre-amplified; the resulting avalanche electrons are transferred into the investigated RWELL, through a transfer gap; the resulting avalanche mimics highly-ionizing events. Events converted within the transfer gap undergo charge multiplication only in the RWELL – mimicking low-ionization events. A scheme for anode-pulses and power-supply currents recording is shown [28].

By varying the injector's gain it was possible to expose the investigated detectors to a broad spectrum of primary-ionization events, spanning from a few tens to several thousand primary electrons. Here the term "primary electrons" (PEs) refers to the electrons that underwent pre-amplification in the injector.

## 3.2 Optical recording: avalanche development studies

An optical readout technique of light emitted from avalanches within THGEM holes permitted studying the avalanche geometrical properties [35]. The optical setup comprised of a lens followed by an intensified CCD camera system - described in detail in [35]. The system allowed for recording a large field-of-view with high spatial resolution; in particular, it permitted studying the gas-avalanche processes within a single hole, e.g. addressing questions like radiation-induced hole-multiplicity, local avalanche asymmetry within a hole etc. The optical-readout investigations were performed in Ne/5%CF$_4$; this mixture provided relatively large avalanche-induced photon-per-electron yields [35].

## 4. Highlights of recent THGEM-structures studies

### 4.1 Pulse shape

Figure 4 shows a comparison between signal shapes of representative double-sided THGEM (without multiplication in the induction gap) and various WELL detectors described in this work. The signals were recorded with an Ortec 125 charge-sensitive preamplifier coupled to the detector's anode, with a digital oscilloscope (Tektronix TDS3052). In the double-sided configuration, the avalanche-induced signal is characterized by a fast rise time of typically a few ns. The WELL signals are slower (in this case ~2 µs rise time for a 0.8-thick single-sided THGEM), due to the avalanche-ion drift within the holes - absent with an induction gap scheme which is sensitive only to the avalanche electrons. Similar signal shapes were measured with all the WELL-structures investigated, including RPWELL. More details can be found in [31].

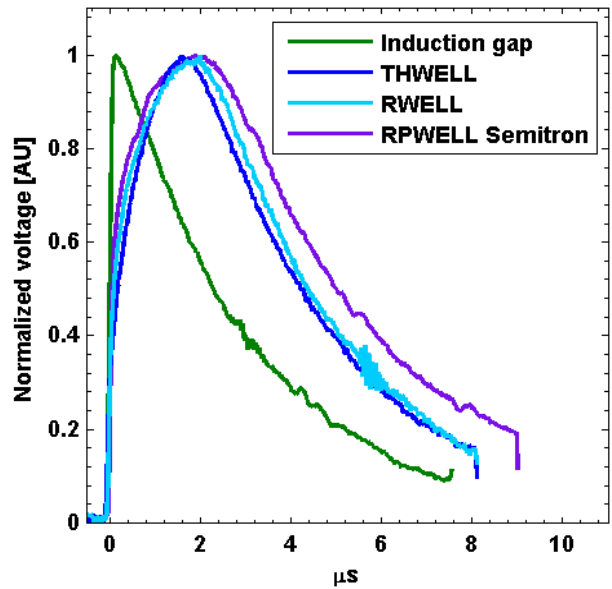

**Figure 4:** Pulse shapes of several THGEM structures, indicated in the legend. All pulses were recorded at a gain of ~6000 in Ne/5%CH$_4$ at atmospheric pressure.

### 4.2 Detector gain, gain saturation and rate dependence

The gain as a function of the voltage applied between the THGEM electrodes (in a double-sided configuration) or between the top electrode and the anode (in WELL configurations) is shown for some representative configurations in Figure 5. Except for the RPWELL, the measurements were stopped with the appearance of occasional discharges.

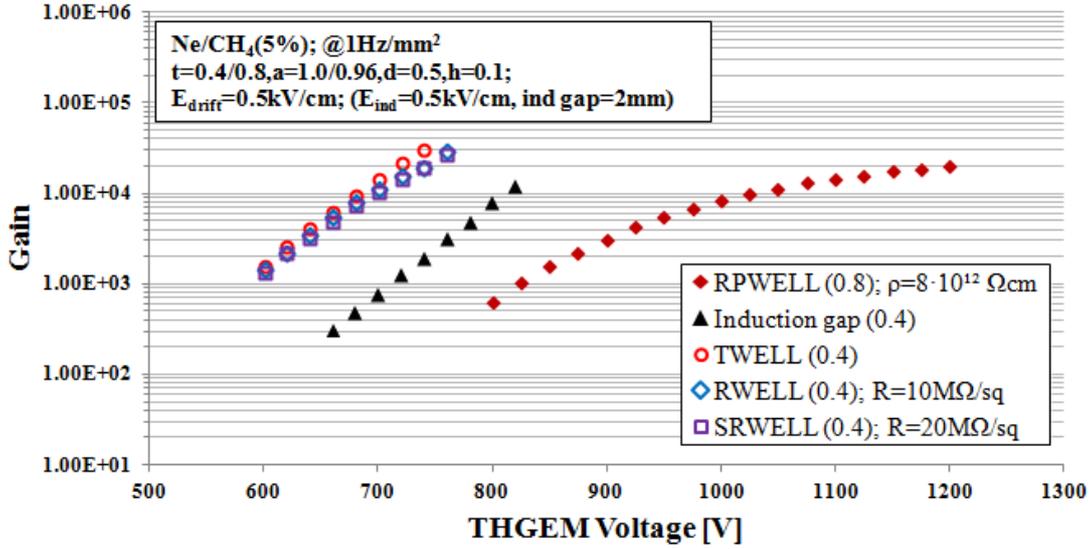

**Figure 5:** The gain as a function of the voltage applied to the THGEM electrodes: 0.4 mm thick double-sided THGEM, 0.4 mm thick THWELL, 0.4 mm thick RWELL with surface resistivity of 10 MΩ/square, 0.4 mm thick SRWELL with surface resistivity of 20 MΩ/square and 0.8 mm thick RPWELL with bulk resistivity of $8 \times 10^{12}$ Ωcm. The THGEM electrode parameters and experimental conditions are given within the figure.

High gains of the order of $10^4$ (with 8 KeV x-ray photons) were reached with all the configurations investigated. At the same operation voltage, the gain of the 0.4 mm thick THWELL and RWELL configurations was higher than that of the 0.4 mm thick double-sided configuration (operated with 2 mm induction gap and $E_{induction}$ = 0.5 kV/cm). This is due to the larger field strength in the closed holes [8]. A unique characteristic was observed with the RPWELL configuration with the highest bulk resistivity (here shown for a 0.8 mm RPWELL coupled to a VERTEC glass anode with a bulk resistivity of $8 \times 10^{12}$ Ωcm [31]: to the best of our knowledge, for the first time, the effect of considerable gain saturation was measured with a THGEM-based detector. Note that since the THGEM electrode was thicker (compared to the other electrodes shown in Figure 5), higher operation voltages were required to reach the same gains. Nevertheless, the operation was stable with no occasional discharges.

The dependence of the gain on the events counting-rate is shown, for similar representative configurations, in Figure 6. The initial gain, at a measured events-rate of ~10 Hz/mm$^2$, was similar for all the configurations (~ 6000). A small rate-dependent gain variation was observed with the 0.4 mm-thick double-sided THGEM configuration, of the order of 10% up to to $10^5$ Hz/cm$^2$. The THWELL, here using a 0.8 mm-thick single-sided electrode, showed a more rapid drop of gain, reaching ~20% at the same rate. Over the same range of event rates, moderate rate-dependent gain-variations of 50% and 30% were observed with the RWELL and the RPWELL respectively (both using 0.8 mm thick electrode); the latter had a 0.6 mm thick Semitron-polymer plate with a bulk resistivity of $2 \times 10^9$ Ωcm [31]. In the highest-resistivity RPWELL configurations the gain dropped rapidly with the rate.

The recorded behaviour of the RPWELLs is consistent with a model in which the clearance times of the avalanche electrons from the bottom of the hole through the resistive plate, are sufficiently long compared to the time difference between consecutive avalanches, resulting in a decreased field and lower gain; in this picture, the higher the resistivity, the longer the clearance times are, and thus the larger is the gain drop at the same rate. The gain drop in the RWELL may be, in principle, explained by a similar argument, here for electron clearance across the resistive layer. The gain drop in the non-resistive structures is likely due to up-charging of the insulating substrate; the differences between the THGEM and THWELL can be attributed to the different field configuration in these structures. The removal of drifting avalanche ions, occurring with typical time of 1-2 μsec, is expected to become important at rates of the order of $10^5$ Hz/mm$^2$, and thus does not play a dominant role in the range of rates studied here.

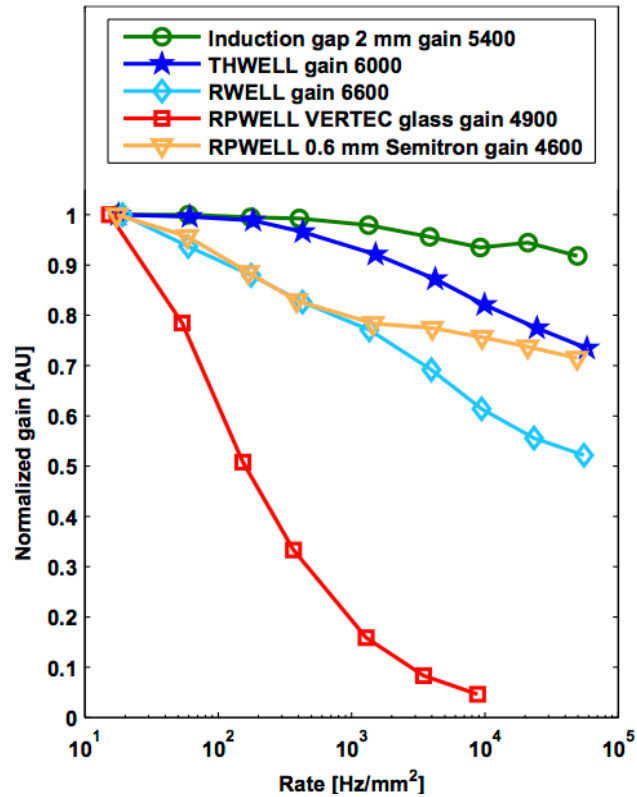

**Figure 6:** Relative detector gain vs. x-rays rate. The different detector configurations, indicated in the figure, were irradiated with a 1 mm diameter 8 keV x-ray beam. The initial gains were of the same order and are indicated in the figure. From [31].

**4.3 Discharge probability and magnitude**

Resistive anodes can effectively quench the energy of discharges. The quenching power of the resistive anode (surface resistivity of 10 MΩ/square) in an RWELL configuration is shown in Figure 7. The charge supplied to the top WELL electrode (calculated from the time integral of the monitoring current) is shown for a THWELL (left) and a RWELL (right). In both configurations, the measurements were conducted with 10 × 10 cm$^2$ WELL electrodes. The typical average charge of 400 nC released in a discharge in the THWELL configuration is quenched to about 40 nC in the RWELL one. In a small fraction of the discharges occurring in the double-sided configuration the measured charge was approximately double than the typical one. Such amount of charge can be due to two consecutive discharges separated but a short time. More details are provided in [36].

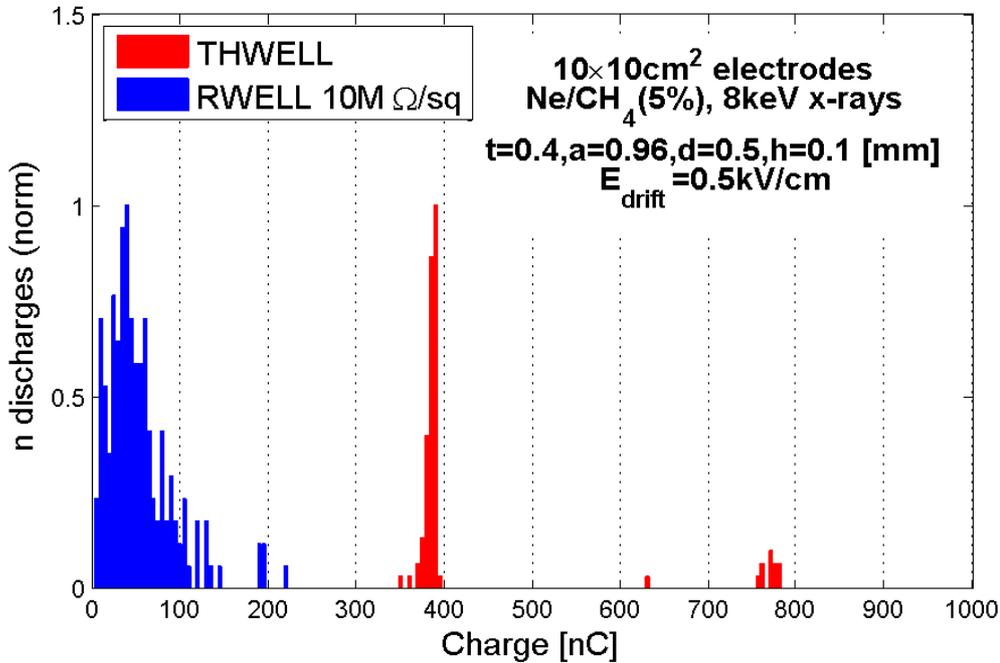

**Figure 7:** The charge supplied to the top WELL electrode (calculated from the time integral of the monitoring current) in a THWELL (red) and a RWELL (blue) structures. From [36].

The discharge probability, defined here as the number of discharges over the number of events, was addressed employing the charge-injector method described above. The discharge probability was measured in conditions close to those expected in a collider experiment. Namely, the investigated detectors were operated at a fixed gain, optimized for the detection of minimum ionizing particles (MIPs). By tuning the injector gain, the investigated detectors were exposed to variable number of primary electrons (PEs) - up to 4 orders of magnitude larger than MIPs, as anticipated by interactions with heavily ionizing particles (HIPs).

An example of a discharge-probability measurement as a function of the number of PEs is shown in Figure 8 for detectors operated at effective gains of ~2500. Two configurations were compared: 1) A double-sided THGEM followed by a 1 mm induction gap, with additional multiplication in this induction gap ($E_{induction}$ = 3 kV/cm), and 2) RWELL configuration (anode surface resistivity of 10 MΩ/square). As illustrated in Figure 8, the dynamic-range (namely the range of primary ionization in which the detector is both efficient and stable) of the double-sided structures is broader than that of the RWELL-based one. Similar conclusions were reached with different double-sided and WELL-based configurations. Indeed, in a double-sided configuration part of the avalanche develops outside of the hole, in the induction gap, where the field is smaller and the physical volume is larger (compared to that inside a hole). These two effects could lead to a higher effective Reather limits. A detailed description of the measurements and a further discussion are given in [28].

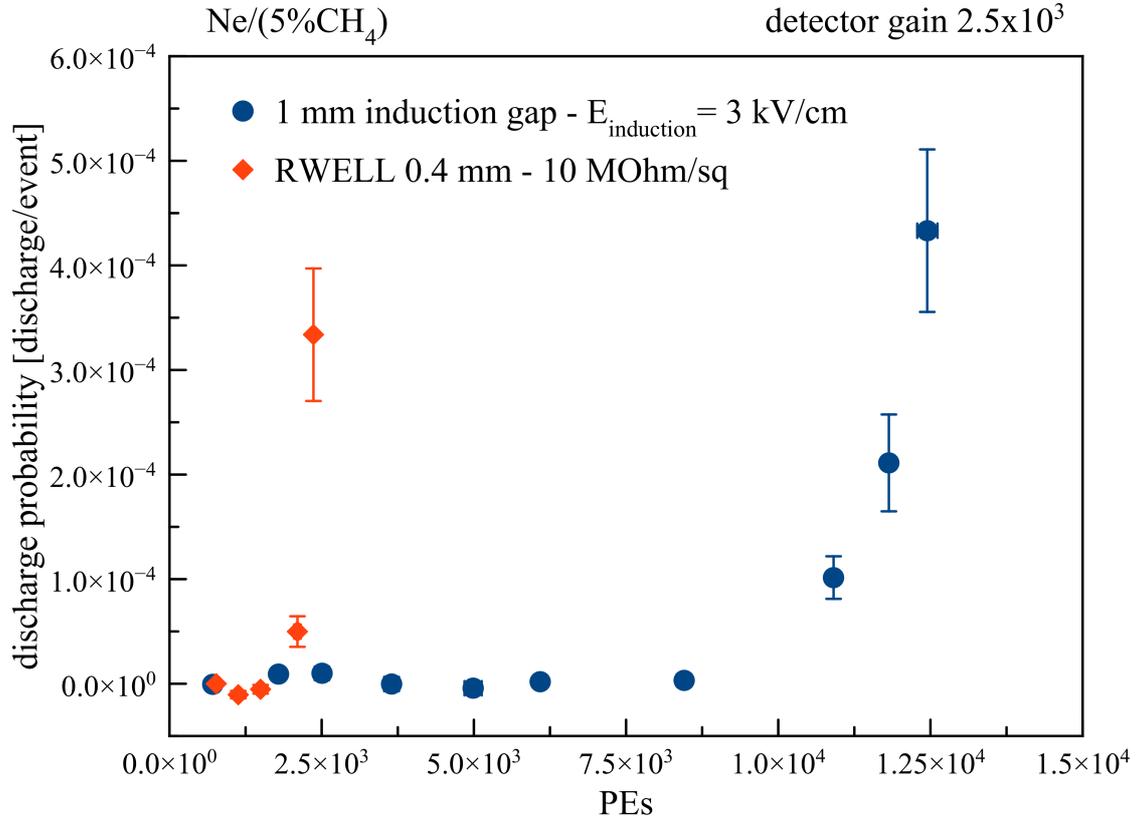

**Figure 8:** Discharge probability as a function of the number of primary electrons (PEs) measured in a 0.4 mm thick THGEM with additional electron multiplication in the induction gap and in a 0.4 mm thick RWELL with surface resistivity of 10 MΩ/square. Total detector gains of 2500, in Ne/5%CH$_4$. From [28].

The response of the RPWELL configuration to events with large primary ionization was found to be significantly different. Even at very high injector gains, of several hundreds, no discharges occurred with most of the resistive materials investigated [31]. However, at injector gains above ∼10$^2$, leakage currents were observed (∼ 10 − 50 nA) which depended on the injector gain; they vanished when the gain was reduced. Even at low incoming particle flux, high injector gains caused detector-gain drops; reducing the injector gain restored the original values [31]. These measurements show that the RPWELL structures are very stable and could be suitable for applications requiring moderate event rate capabilities.

### 4.4 Avalanche characteristics

Using the optical setup described above [35] we measured the hole-multiplicity (number of holes sharing event-induced avalanches) in single- and in multi-stage double-sided THGEM configurations. Figure 9 shows typical x-ray induced single-event avalanches. As expected from electron diffusion and photon-induced avalanche broadening, the hole-multiplicity increased with the number of multiplication stages, with larger transfer gaps between the consecutive stages and under lower transfer fields.

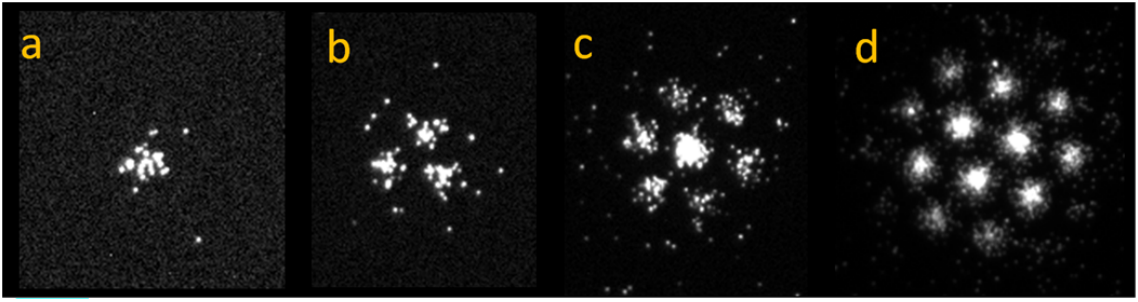

**Figure 9:** Examples of single-event avalanches, induced by 5.9 keV x-rays, recorded in different detector configurations. a) Single-THGEM with a reversed drift field at gain ~$10^4$; b) Single-THGEM ($E_{drift}$=0.5 kV/cm, gain ~$10^4$); c) Double-THGEM with 8 mm transfer gap ($E_{drift}$=0.5 kV/cm, $E_{trans}$=0.5 kV/cm, gain ~$5\times10^5$); d) Triple-THGEM with 8 mm and 10 mm transfer gaps ($E_{drift}$=0.5 kV/cm, $E_{trans}$=0.5kV/cm, gain ~$10^7$). The images are unprocessed, but the contrast has been adjusted to improve visibility. Gas: Ne/5%$CF_4$. From [35].

The optical method also permitted exploring the avalanche formation within the THGEM holes; one of the observations was that the avalanches are asymmetric - developing away from the hole-edge, in locations correlated with the incoming particle's position. This asymmetrical development of the avalanche could be used to improve position resolution in THGEM detectors. For more details see [35].

## 5. Summary and discussion

The THGEM is a robust electrode, industrially manufactured over large areas by standard printed-circuit board techniques. We reviewed here some new THGEM-detector concepts and the properties of various THGEM-like structures; the focus was on those of WELL-based configurations derived from closed-bottom THGEM electrodes: the Thick WELL (THWELL), Resistive WELL (RWELL), Segmented Resistive WELL (SRWELL) and Resistive-Plate WELL (RPWELL). The fast rise-time of the signals in these WELL-based structures is followed by a slowly rising ion component. Gains of order several $10^4$ were reached with soft x-rays, also in single-stage configurations. A clear indication for gain saturation was seen, for the first time, in the RPWELL with high (~$10^{12}$ $\Omega$cm) bulk resistivity. The saturation is most likely due to slow transport of the avalanche electrons through the high resistivity anode, resulting in charge accumulated at the bottom of the hole and decreasing the strength of the electric field inside the hole.

The dependence of the gain on the recorded event-rate varied among the configurations investigated; in the RPWELL it strongly depended on the resistivity of the anode. While the gain drop of the double-sided THGEM and single-side THWELL configurations was lower than 10% over event rates in the range of 10 to $10^5$ Hz/mm$^2$, gain drops of order 30% were measured for configurations with moderate resistivity values over the same range. The gain of configurations with very high bulk resistivity dropped very fast. This could also be explained by the degradation of the electric field due to the charge accumulated at the bottom of the holes.

Two methods were proposed to investigate gaseous-detector properties. The first one is the charge-injector method, allowing for evaluating, in the laboratory, the detector's response over a broad dynamic range - mimicking well-defined highly ionizing events. The second is an optical method, providing a precise tool for investigating the geometrical properties of charge avalanches by recording avalanche photons. While the potential of these methods was demonstrated in THGEM-structures, both could be useful for systematic studies of other gaseous radiation detectors.

The response of the different structures was investigated over a broad ionization dynamic-range exploiting the injector method. The dynamic range of double-sided configurations operated with moderate additional multiplication in the induction gap following a THGEM was found broader than that of the THWELL and the RWELL. This observation is consistent with the known avalanche-development model suggesting that in a double-sided configuration part of the avalanche develops outside of the THGEM hole (the avalanche extension depends on the respective fields in the hole and in the induction gap), effectively resulting in a higher Reather limit. Some of the RPWELL configurations were operated in stable conditions also in presence of very large primary ionization, possibly due to charge saturation.

The optical setup permitted evaluating the hole-multiplicity in single events recorded in THGEM configurations. It was shown that the average number of "illuminated" holes per event depends on the detector's configuration; the study further revealed the asymmetric nature of the avalanche, even within a single hole – being correlated with the position of the incoming particles.

The applicative potential of THGEM-based gaseous multipliers is large and these robust detectors are already being considered for a variety of applications discussed in the introduction. We have shown a selection of WELL-like multipliers, with metal and resistive anodes; the choice of the appropriate structure should be made and optimized for a specific application and operation conditions. A most promising novel structure is the RPWELL; equipped with a 0.6 mm thick Semitron-polymer resistive plate, it provided so far the best performance in terms of stable operation over a broad dynamic range, even when exposed to highly ionizing particles and at rather high particle flux. The research efforts on the WELL structures focus towards operation stability, optimization of the resistive materials, localization properties and studies of large-area detectors.

## Acknowledgments

This work was supported in part by the Israel-USA Binational Science Foundation (Grant 2008246) and by the I-CORE Program of the Planning and Budgeting Committee and The Israel Science Foundation (grant NO 1937/12). A. Breskin is the W.P. Reuther Professor of Research in the Peaceful use of Atomic Energy.